# Rotationally Induced Surface Slope-Instabilities and the Activation of CO$_2$ Activity on Comet 103P/Hartley 2

Jordan K. Steckloff[1], Kevin Graves[2], Toshi Hirabayashi[2,3], H. Jay Melosh[1,2], James Richardson[4]

[1] Purdue University, Department of Astronomy & Physics, West Lafayette, IN
[2] Purdue University, Department of Earth, Atmospheric, & Planetary Sciences, West Lafayette, IN
[3] University of Colorado, Boulder CO
[4] Arecibo Observatory, Arecibo, Puerto Rico

Abstract

Comet 103P/Hartley 2 has diurnally controlled, CO$_2$-driven activity on the tip of the small lobe of its bilobate nucleus. Such activity is unique among the comet nuclei visited by spacecraft, and suggests that CO$_2$ ice is very near the surface, which is inconsistent with our expectations of an object that thermophysically evolved for ~45 million years prior to entering the Jupiter Family of comets. Here we explain this pattern of activity by showing that a very plausible recent episode of rapid rotation (rotation period of ~11 [10-13] hours) would have induced avalanches in Hartley 2's currently active regions that excavated down to CO$_2$-rich ices and activated the small lobe of the nucleus. At Hartley 2's current rate of spindown about its principal axis, the nucleus would have been spinning fast enough to induce avalanches ~3-4 orbits prior to the DIXI flyby (~1984-1991). This coincides with Hartley 2's discovery in 1986, and implies that the initiation of CO$_2$ activity facilitated the comet's discovery. During the avalanches, the sliding material would either be lofted off the surface by gas activity, or possibly gained enough momentum moving downhill (toward the tip of the small lobe) to slide off the tip of the small lobe. Much of this material would have failed to reach escape velocity, and would reimpact the nucleus, forming debris deposits. The similar size frequency distribution of the mounds observed on the surface of Hartley 2 and chunks of material in its inner coma suggest that the 20-40 meter mounds observed by the DIXI mission on the surface of Hartley 2 are potentially these fallback debris deposits. As the nucleus spun down (rotation period increased) from a period of ~11 hours to 18.34 hours at the time of the DIXI flyby, the location of potential minima, where materials preferentially settle, migrated about the surface, allowing us to place relative ages on most of the terrains on the imaged portion of the nucleus.

Introduction

The DIXI (Deep Impact eXtended Investigation) flyby of comet 103P/Hartley 2 on November 4, 2010, revealed the nucleus to be a small, bilobate, but highly active world (A'Hearn et al. 2011). Hartley 2's observed activity was dominated by a region of vigorous $CO_2$ sublimation at the tip of its small lobe, which was illuminated during DIXI's closest approach to the nucleus (A'Hearn et al. 2011). This region's $CO_2$ production was significantly diminished half a rotation later while unilluminated (A'Hearn et al. 2011), suggesting that it is diurnally controlled, yet it still dominated the $CO_2$ production from the illuminated large lobe (Feaga et al. 2014). While bilobate comet nuclei are fairly common (Keller et al. 1986; Oberst et al. 2004; Harmon et al. 2010; A'Hearn et al. 2011; Sierks et al. 2015), diurnal control of $CO_2$ sublimation had never before been observed by spacecraft in situ.

Diurnal control requires that Hartley 2's $CO_2$ ices are located within a few diurnal thermal skin depths of the nucleus, which extends no more than a few centimeters below the surface. However, this is inconsistent with the expected thermophysical evolution of a Jupiter Family Comet (JFC) like Hartley 2, which spend typically ~45 million years as a Centaur object (Duncan et al. 2004) before migrating into the Jupiter family. In the Centaur region of the Solar System carbon monoxide and carbon dioxide ices sublimate vigorously enough to drive cometary activity in this region of space (Sekanina, 1992; Steckloff & Jacobson, 2016). Volatile sublimation over such a long dynamical lifetime should produce a chemical stratification of the surface layers of the nucleus, with more volatile ices receding into the interior of the nucleus, while less volatile ices such as water ice remain closer to the surface. Observations by the Deep Impact spacecraft suggest that the diurnally controlled $H_2O$ ices of comet Tempel 1 receded less than ~1-10 cm below the surface (Groussin et al. 2007; Davidsson et al. 2013). If we scale this diurnal skin depth to the seasonal (orbital) thermal skin depth, then the more volatile and seasonally active $CO_2$ ices (Feaga et al. 2007) were located within ~10 – 100 cm of the surface. This is consistent with the lack of significant variation in the $CO_2/H_2O$ ratio in the Deep Impact ejecta versus Tempel 1's preimpact ambient outgassing (A'Hearn et al. 2005), which should be similar if the difference in the depths of the $CO_2$ and $H_2O$ sublimation fronts below the surface is small in relation to the ~10 m (Richardson & Melosh, 2013) or ~50 m (Schultz et al. 2013) depth of the Deep Impact crater. Thus, the expectation that near-surface $H_2O$ sublimation drives the diurnal activity of a JFC is consistent with high-resolution spacecraft observations of JFCs (Feaga et al. 2007; Gulkis et al. 2015; Sierks et al. 2015). Comet 67P/Churyumov-Gerasimenko, like Hartley 2, also exhibits diurnal control of $CO_2$ sublimation (Hässig et al. 2015).

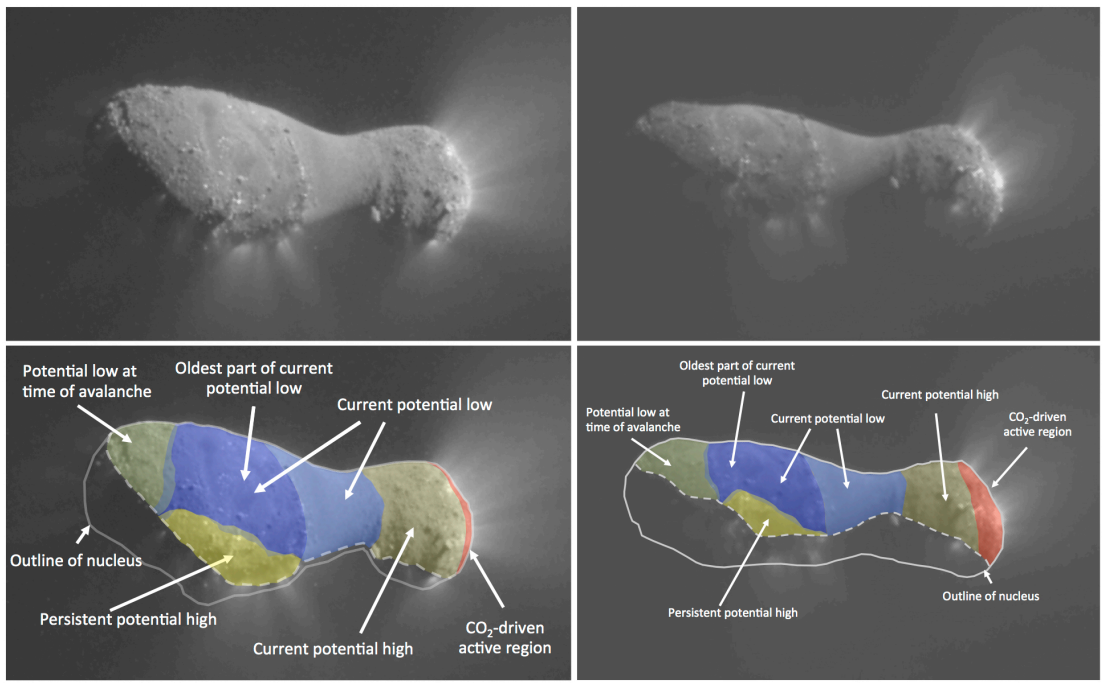

**Figure 1:** *Overview of relevant regions of comet 103P/Hartley 2.* The nucleus of comet Hartley 2 exhibits various terrain types visible in these original and annotated images from the DIXI flyby (MRI-VIS frame 5004052 (*left*) and frame 5004064 (*right*)). These two images cover nearly the entire imaged surface of the nucleus of comet Hartley 2. Although the Deep Impact spacecraft only obliquely imaged the CO2-driven active region, the entire tip of the small lobe is dominated by $CO_2$ activity. The rest of the nucleus surface is a combination of knobby, hummocky terrain and smooth regions. As a result of this work, we have determined the location of current and former locations of low net potential (gravitational plus rotational potential), which migrate about the surface as the rotation state of the nucleus changes and lengthens to the observed period during the DIXI flyby. If we assume that avalanches and setting of debris into potential lows are responsible for forming the observed surface of Hartley 2, then we can derive relative ages for its observed surface terrains: (oldest to youngest): persistent potential high, $CO_2$-driven region, potential low at time of avalanche (which remained a potential low until the rotation period exceeded 14 hours), oldest part of current potential low, and current potential low.

For Hartley 2 to exhibit diurnally controlled $CO_2$ sublimation, Hartley 2 must either have had an unusually short migration through the Centaur region into the Jupiter Family that limited the thermophysical evolution of the surface layers of the nucleus, or some mechanism must have recently removed the thermally evolved surface layers that are expected to overlie $CO_2$-rich layers. Here we show that an episode of fast rotation in the recent past of Hartley 2 could have generated surface slope instabilities that exposed buried $CO_2$-rich ices in the region of the observed activity.

At the time of the Deep Impact flyby, the nucleus of Hartley 2 was in a tumbling rotation state (A'Hearn et al. 2011; Knight & Schleicher, 2011; Samarasinha et al. 2011), with a rotation period about its principal axis of 18.3 hours (A'hearn et al. 2011; Drahus et al. 2011) and rotation period about its long axis of 27.79 hours (A'Hearn et al. 2011). However, sublimative torques can change comet spin periods, and during the Deep Impact flyby, Hartley 2's rotation period about its principal axis was lengthening at an estimated rate of 1.3±0.2 min/day based on DIXI flyby imagery (Belton et al. 2013) or 1.00±0.15 min/day from ground-based observations (Drahus et al. 2011), while its rotation period about the long axis (the rolling motion of the nucleus) first increased before decreasing again during the encounter (Belton et al. 2013). Other ground-based observations are consistent with these rapid changes in the rotation state of the nucleus (Knight & Schleicher, 2011; Meech et al. 2011; Samarasinha et al. 2011; Knight et al. 2015).

This suggests that Hartley 2 was likely spinning much faster about its principal axis in the recent past. At the rate of angular deceleration observed during the DIXI encounter, Hartley 2 would have been spinning fast enough to break apart (disrupt) only ~20 orbits in the past (Drahus et al. 2011). However, this assumes that the nucleus would survive to the disruption limit of a gravity-dominated ellipsoid (Pravec & Harris, 2000). We used Finite Element Model (FEM) analysis (Hirabayashi & Scheeres, 2015) to consider the rotationally-induced stresses that concentrate at the waist of Hartley 2's bilobate structure, and found that its nucleus would fission into two lobes at rotation periods shorter than ~8 hours, assuming a material tensile strength on the order of ~1 Pa (Sekanina & Yeomans, 1985; Asphaug & Benz, 1996). Thus, Hartley 2 would have been rotating at its disruption limit only ~10 orbits in the past. This suggests that the $CO_2$ activity driving Hartley 2's sublimative torques is either less than ~70 years old, or that its nucleus migrated unusually quickly through the Centaur region, such that the nucleus did not significantly thermophysically evolve.

In the case of rapid migration, Hartley 2's orbit would need to have evolved from a perihelion outside of ~10 AU into the Jupiter Family of comets in a timespan on the order of centuries or shorter for its $CO_2$ activity to remain diurnally controlled and the nucleus to remain intact. However, this is exceedingly unlikely when compared to the typical migration timescale through the centaur region of ~45 million years (Duncan et al. 2004). Even if we assume that Hartley 2 migrated through the Centaur region in negligible time and was directly injected into the Jupiter Family, the median dynamical lifetime of a JFC is ~325,000 years (Duncan et al. 2004) and $CO_2$ sublimates vigorously throughout the JFC region of space (Steckloff et al. 2015; Steckloff & Jacobson, 2016), making it exceedingly unlikely (~0.02%) that Hartley 2 has been in the Jupiter Family for less than this ~70 year maximum age of activity.

Instead, we investigate the alternative, that comet Hartley 2's $CO_2$ activity is young, and thus was a relatively dormant comet that was reactivated in the recent past. We propose that rotationally induced avalanches preferentially exposed buried $CO_2$-rich materials on the surface of Hartley 2's small lobe, where solar radiation could diurnally control $CO_2$ sublimation, generating the observed activity of the nucleus. We explore this case numerically.

Methods

We explore the effects of spin rate changes on the surface of Hartley 2's nucleus by computing the stability of its slopes at rotation periods between four hours (where the nucleus is unstable and breaks apart) and its rotation period of 18.34 hours at the time of the DIXI flyby. Although Hartley 2 is in a tumbling rotation state (A'Hearn et al. 2011; Knight & Schleicher, 2011; Samarasinha et al. 2011), we initially ignore the slower rotation about its long axis (period of 27.79 hours at the time of the flyby [A'Hearn et al. 2011]), and assume principal axis rotation.

We assume that avalanches, which remove surface materials while leaving underlying materials undisturbed, are responsible for exposing $CO_2$-rich ices on the surface of the nucleus. Avalanches occur when the slope angle of a surface exceeds its angle of repose (Lambe & Whitman, 1969). We compute the surface slope angles of Hartley 2 by first computing the net acceleration vector (the sum of the gravitational and centrifugal acceleration) at the center of each facet of the Thomas et al. (2013) shape model of Hartley 2 using the GRAVMAP code, which is based on the method of Werner (1994). We then compute the angle between the net acceleration vector and the vector normal to the facet of the shape model to obtain the surface slope angle of the facet. We next identify facets of the shape model with slope angles that exceed the angle of repose of Hartley 2's regolith, an unstable condition that precedes landslides and avalanches.

Critical to this method is the density of the nucleus, which is assumed to be uniform. The high relative encounter velocity of the Deep Impact spacecraft with Hartley 2's small nucleus prevented a direct measurement of its density from gravitational deflection of the spacecraft (A'Hearn et al. 2011). However, A'Hearn et al. (2011) considered that the smooth waist of the nucleus is likely a ponded depositional feature, which requires a density of at least 220 kg/m³ for this region to occupy a gravitational low. Richardson & Bowling (2014) considered that the waist is likely the result of a fluidized flow, and would therefore approximate an equipotential surface. By minimizing the variance of the effective potential (gravitational plus rotational) for the observed portion of the waist at the principal rotation period during the Deep Impact flyby, they estimated the nucleus density to be $\rho$= 200 (140-520) kg/m³. Thomas et al. (2013) further considered the changing rotation state of the nucleus to refine this estimate to $\rho$= 300 (200-400) kg/m³.

The angle of repose ($\alpha$) of a surface (the maximum stable slope angle) depends on the angle of internal friction ($\phi$), pore pressure of fluids within the material ($p_{pore}$), material density ($\rho$), cohesive strength ($\sigma_c$), and the local gravity field (**g**) of the body

$$\sigma_S = \sigma_c + (\sigma_n - p_{pore}) \tan \phi \qquad (1)$$

$$\sigma_S = \rho g h \sin \alpha \qquad (2)$$

$$\sigma_n = \rho g h \cos \alpha \qquad (3)$$

where $\sigma_S$ and $\sigma_n$ are, respectively, the shear stress and normal stress exerted on the surface by a surface block or layer, and $h$ is the thickness of the unstable layer (Melosh, 2011). The angles of internal friction for geologic materials are remarkably uniform, typically ~30°-45° (Lambe &

Whitman, 1969). We conservatively assume that pore pressure ($p_{pore}$) within the regolith of Hartley 2 is negligible, which will result in more stable surfaces and higher angles of repose.

We also assume that the surface regolith of Hartley 2 is non-cohesive, based on the presence of the smooth waist of Hartley 2 that is believed to be a flow deposit that fluidized by $H_2O$ sublimation (A'Hearn et al. 2011; Thomas et al. 2013; Richardson & Bowling, 2014). Because this regolith would fail to fluidize if its cohesive strength were greater than the vapor pressure of the sublimating $H_2O$, the vapor pressure of $H_2O$ provides a maximum constraint to the cohesive strength of the regolith. If $H_2O$ ice were located right at the surface of the smooth waist (as opposed to mixed within it), its vapor pressure at perihelion with the Sun at the zenith would be ~0.1 Pa (Steckloff et al. 2015; Steckloff & Jacobson, 2016), however this constraint would only allow for the onset of fluidization for a very brief moment of Hartley 2's orbit. If we assume that the waist is able to fluidize for a few months preceding perihelion, then the constraint on the cohesive strength of the regolith drops by nearly an order of magnitude. The constraint on the cohesive strength drops even further if we assume that the sublimating $H_2O$ ice deposits are located within the regolith (rather than on top of it). While the vapor pressure briefly increases as one covers the ice with a few layers of dust particles (Blum et al. 2014), the low thermal inertial of cometary regolith inhibits the amount of heat that reaches sublimating ice below the surface, and results in a lower vapor pressure that would require an even weaker regolith to form a fluidized flow. It is therefore reasonable to assume that the regolith of Hartley 2 has such weak cohesion, that it effectively behaves as a non-cohesive material.

In this case of regolith without significant pore pressure or cohesion, the angle of repose ($\alpha$) in equations (2)-(3) becomes equal to the angle of internal friction ($\phi$), and is therefore expected to lie between 30°-45°. This is consistent with the frequency distribution of the gravitational slope angle of terrains on the nucleus of comet 67P/Churyumov-Gerasimenko, which drops off rapidly as gravitational slope angles exceed 25°-45° (Groussin et al. 2015). We therefore conservatively choose to only consider slopes less than 30° to be stable, and slopes exceeding 45° to be unstable. It is unclear if surface slopes between 30° and 45° are stable or unstable without a more thorough understanding of the structural properties of Hartley 2's regolith. We vary the principal axis rotation period of Hartley 2 in 1 hour increments, compute the resulting surface slope angles, and identify unstable regions of the nucleus surface for each rotation period.

Results

We find that the surface of comet Hartley 2 is generally very stable during the DIXI flyby (spin period of 18.34 hours), with all the surface (except for a single scarp on the large lobe) possessing a surface slope angle less than 20°, and therefore stable. The surface slopes of the large lobe's scarp are maximally between 30° and 45°, leaving unclear the stability of its surface. However, this feature is associated with a dust jet (Bruck Syal et al. 2013), and so it is plausible that this scarp is unstable or was unstable in the recent past.

As we spin up the nucleus of Hartley 2 to a rotation period of 13 hours, a ring of terrain outlining the $CO_2$-driven activity of the small lobe steepens to surface slope angles of 30° to 45° degrees (see Figure 2). While this does not necessarily indicate that the tip of the small lobe becomes unstable and prone to avalanches, it does suggest that the tip of the small lobe is trending toward slope instability as the nucleus is spun up. Interestingly, the rest of the nucleus outside of the scarp on the large lobe remains relatively flat, with surface slopes less than ~20°. At a 13 hour rotation period, the scarp on the large lobe steepens to a slope angle of ~45° indicating that it is likely unstable without cohesion.

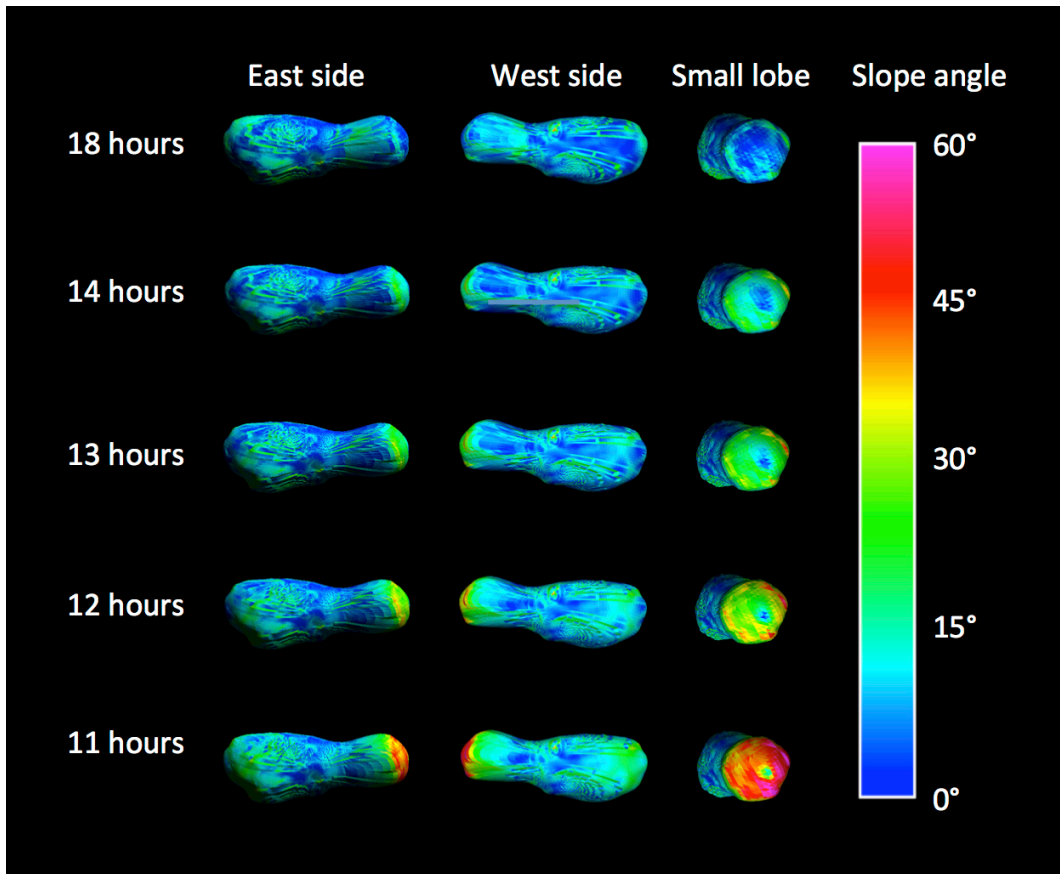

**Figure 2**: *Surface slopes of 103P/Hartley 2 under different principal axis rotation rates.* As the nucleus of comet 103P/Hartley 2 is spun up from it's DIXI flyby rotation period of 18.34 hours, the slopes on the tip of the small lobe of the nucleus increase significantly more than any other place on the nucleus, and become highly unstable (above 45°) at a rotation period of 11 hours. This suggests that, at an 11 hour rotation period, avalanches would set in on the tip of the small lobe, excavating buried $CO_2$ ices and activating this region of the nucleus. Interestingly, the rest of the nucleus remains at roughly the same slope, regardless of rotation period, and would not experience avalanches. The East Side view of the nucleus is similar to the spacecraft view of the nucleus in Figure 1.

At a rotation period of 11 hours, the surface slopes of the source region of the $CO_2$-driven activity at the tip of the small lobe exceed 45° and become clearly unstable. Under these conditions, avalanches will excavate buried materials, and surface materials will flow downhill, toward the tip of the small lobe (see Figure 3). Interestingly, these avalanches are almost exclusively restricted to the active region of the small lobe, as the rest of the nucleus outside of the scarp on the large lobe and a few isolated facets remains stable with surface slope angles less than 20°. The remarkable geographic correlation between surface slope instabilities and the $CO_2$-driven active terrains on the small lobe of Hartley 2 strongly suggest that the nucleus recently experienced an episode of fast rotation (with a period of ~11 hours), which excavated buried $CO_2$ ices to the surface of the nucleus where they currently drive activity.

This result is robust against uncertainties in the bulk density of the nucleus. While our previous computations assume a nucleus bulk density ($\rho$) of 300 kg/m³ (Thomas et al. 2013), a decrease in the nucleus bulk density makes the surface slopes more prone to change under rotational spin up, while an increase in density makes the surface slopes more resistant to change. As a result, when we run the same analysis and vary the nucleus bulk density ($\rho$), we find that surface slope instabilities set in at the tip of the small lobe at a rotation period of ~13 hours for a bulk nucleus density of 200 kg/m$^3$, and at a rotation period of ~10 hours for a bulk nucleus density of 400 kg/m$^3$. However, the distribution of surface slopes is effectively unchanged under these differing densities, and the same regions and isolated facets of the nucleus shape model remain stable/unstable across the uncertainty in the nucleus bulk density.

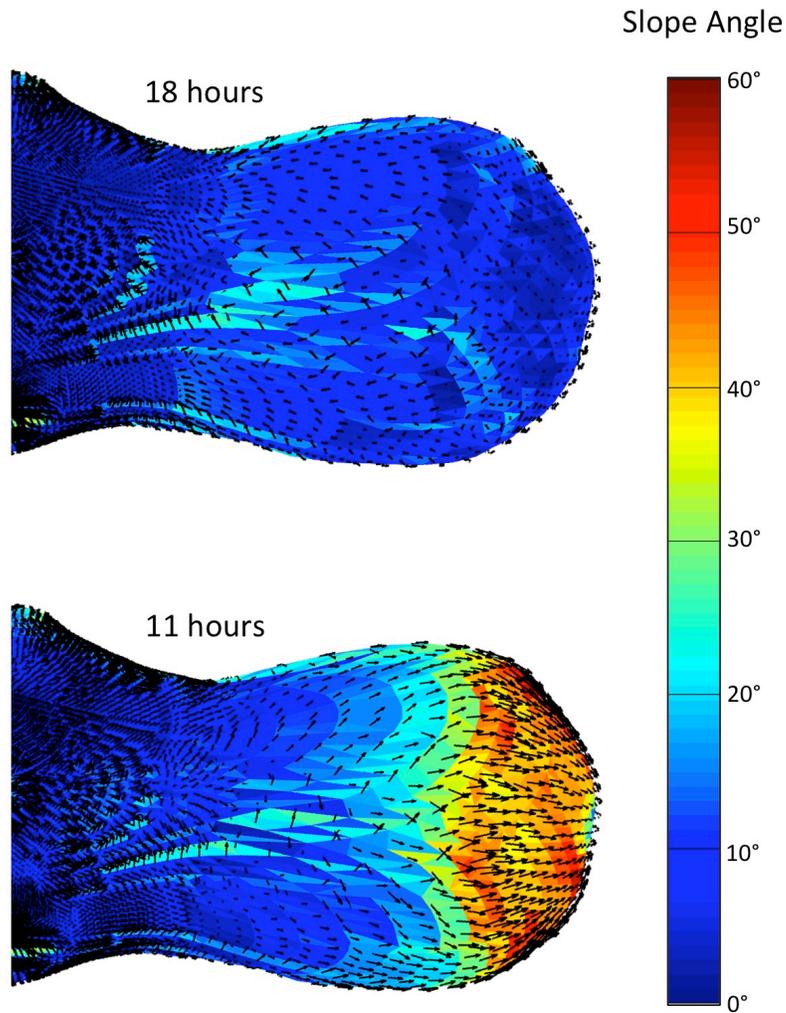

**Figure 3:** *A plot of the downslope directions of Hartley 2's small lobe.* The slope angles and direction are highly dependent on the rotation state of Hartley 2's nucleus. This figure shows the slope angles (by color) and downslope directions (black arrows) of each facet of the nucleus on and near the small lobe for two important rotation states: the 18 hour principle rotation period during the EPOXI flyby of the nucleus, and the 11 hour rotation period that corresponds to the onset of surface slope instabilities of the small lobe. While the downslope direction of the small lobe during the EPOXI flyby was generally toward the waist of the nucleus, during the onset of surface slope instabilities, the slope direction was toward the tip of the lobe.

We also consider the non-principal axis rotation (rotation about the long axis) of the nucleus. We assume that the nucleus is not rotating about its principal axis, and find that non-principal axis rotation leads to surface-slope instabilities when its period is shorter than ~7-8 hours. However, at a period of 6 hours, non-principal axis rotation starts to break up the nucleus, providing a hard constraint on the rotation history of the nucleus (see Figure 4). There is therefore a narrow window of rotation states (period of 6-8 hours) in which non-principal axis rotation can affect our results without breaking up the nucleus. These results are robust against the various principal axis rotation states studied. Furthermore, these rotation states are highly unstable and would require strong sublimative torques to prevent the nucleus from reorienting into a principal axis rotation, which weaken rapidly as the comet recedes from the Sun after perihelion. It is therefore unlikely that non-principal axis rotation has played a role in resurfacing Hartley 2's nucleus.

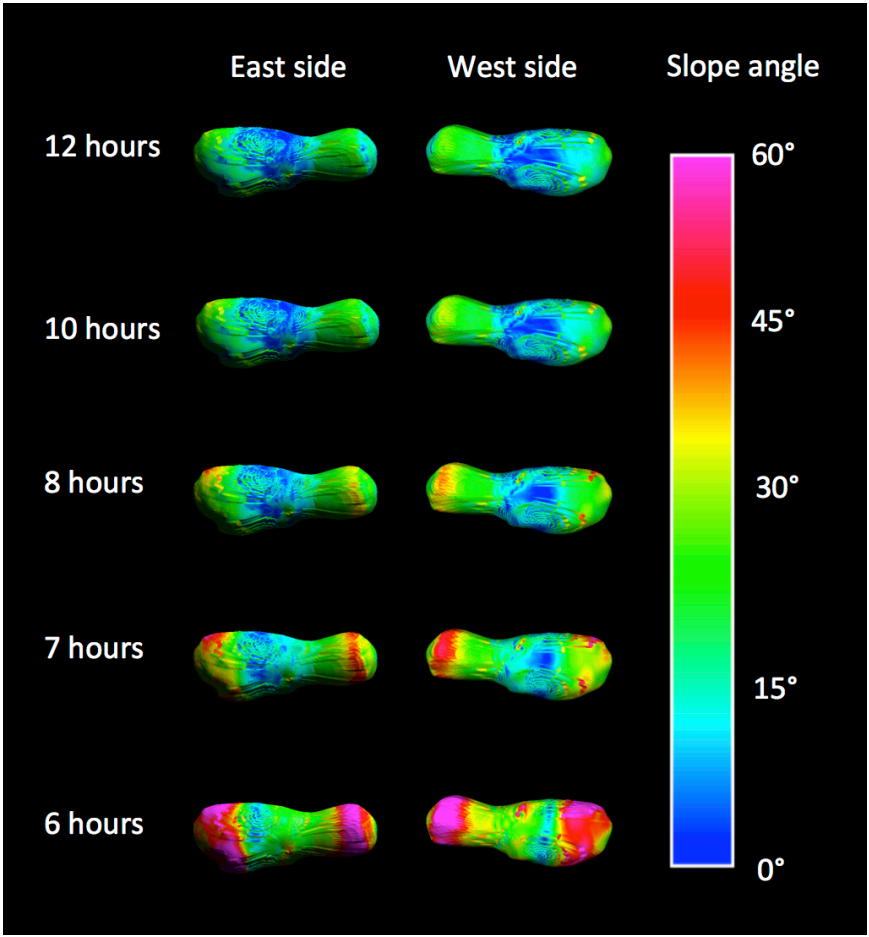

**Figure 4**: *Surface slopes of 103P/Hartley 2 under different non-principal axis rotation rates.* We compute the effects of non-principal axis rotation about the long-axis of the nucleus on the surface slope distribution of Hartley 2. As the nucleus of Hartley 2 is spun up about this axis, The slopes farthest from the long-axis of the nucleus steepen. At non-principal axis rotation periods shorter than 8 hours, the lobes of the nucleus begin to experience significant surface slope instabilities that could affect our results. However, at non-principal axis rotation periods shorter than ~6 hours, these unstable regions start to break apart from the nucleus. Because these regions were observed to be intact, this provides a hard constraint on the recent non-principal axis rotation period of Hartley 2.

Although we assume the regolith of the comet to be cohesionless, the underlying, non-thermally evolved materials likely possess significant strength. We use finite element model (FEM) analysis (Hirabayashi & Scheeres, 2015) to compute the internal stress state of Hartley 2's nucleus and determine the minimum strength required to maintain the interior structural stability of the nucleus at the onset of surface slope instabilities on the small lobe. We input the Thomas et al. (2013) shape model into ANSYS FEM software and assume that the nucleus deforms plastically. At the DIXI flyby rotation period of 18.34 hours, the nucleus experiences entirely compressional stresses, but does not fail compressionally (structurally stable). However, at the onset of surface slope instabilities, the waist of Hartley 2 enters a tensile state (while the two lobes of the nucleus remain compressional), and requires a tensile strength of at least 2.8 (2.0 - 4.0) Pa to remain structurally stable (see Figure 5).

We compare this to known bulk strength estimates of comet nuclei and find that it is consistent with the constraints on the tensile strength of comets Shoemaker-Levy 9 (D/1993 F2) and Brooks 2 (16P) of <6.5 Pa (Asphaug & Benz, 1996) and <2 Pa (Sekanina & Yeomans, 1985) respectively. This constraint is also consistent with estimates of comet Churyumov-Gerasimenko's (67P's) cohesive strength of 1-16 Pa (Bowling et al. 2015) and tensile strength of 3-15 Pa (Groussin et al. 2015) and 20 Pa (Thomas et al. 2015). It is also consistent with the ~1 Pa order unconfined crushing strength of Comet ISON (C/2012 S1) (Steckloff et al. 2015) and >17 Pa cohesive strength of Comet Wild 2 (81P) (Melosh 2011; Steckloff et al. 2015).

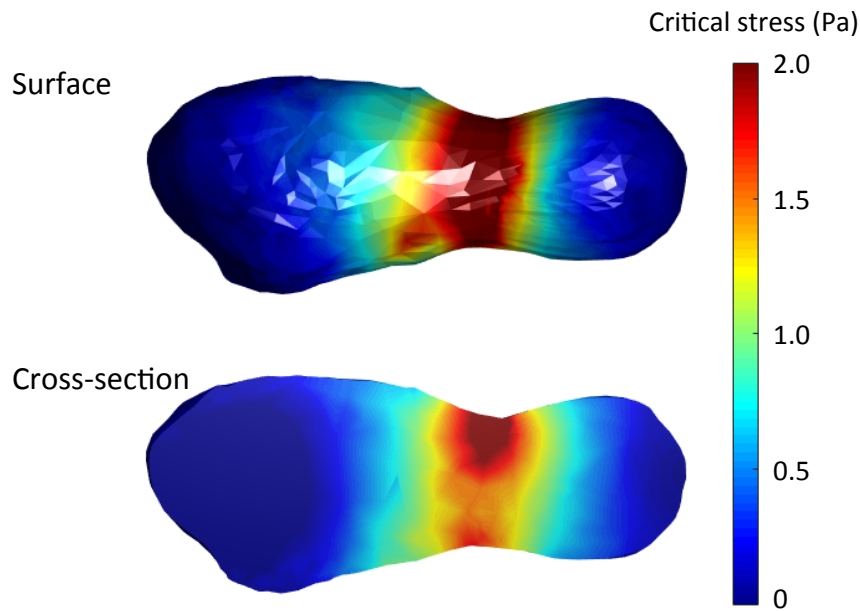

**Figure 5:** *Finite Element Model (FEM) Analysis of Plastic Stresses in the body of Hartley 2 at Critical Rotation Period.* We model the internal stress state of Hartley 2's nucleus at the critical rotation period at which surface slope instabilities and their resulting avalanches set in on the small lobe (11 hour period for a bulk density of 300 kg/m$^3$). We assume plastic deformation of the nucleus material, and compute the highest stresses experienced by the nucleus material. Although we assume that a non-cohesive regolith covers the nucleus, the interior of the nucleus must have ~2.8 Pa of cohesive strength to prevent the nucleus from breaking, consistent with strengths of other cometary nuclei (Sekanina & Yeomans, 1985; Asphaug & Benz, 1996; Melosh, 2011; Bowling et al. 2014; Groussin et al. 2015; Thomas et al. 2015; Steckloff et al. 2015).

Discussion

Small, isolated jets appear on the large lobe of the nucleus (Bruck Syal et al. 2013), and it is conceivable that the scattered, isolated facets of the large lobe that experience surface slope instabilities at faster spin rates may be associated with jets. However, the contribution of these jets to the overall production of the nucleus is likely very minor, otherwise the light curve amplitude during DIXI's approach to Hartley 2 (A'Hearn et al. 2011) would not be as dramatic. Additionally, these unstable facets were on the unilluminated side of the nucleus during the DIXI flyby. The shape of this unilluminated region was determined from the silhouette of the nucleus against light scattered by the dust coma, which introduces uncertainties in the shape of the nucleus that are three times greater than the method of tracking surface features on the illuminated part of the nucleus (Thomas et al. 2013). It is therefore possible that the surface slope instabilities of these facets are artifacts of imperfections in the Thomas et al. (2013) shape model. Furthermore, although the sources of some of the small jets are associated with geologic features of the nucleus (Bruck Syal et al. 2013), these source areas are smaller than, or comparable to, the 2 degree (~10-44 m) resolution of the Thomas et al. (2013) shape model. Because our method is limited by the resolution of the shape model, studying the activation of these isolated jets, or the effects of isolated unstable facets of the shape model, is difficult and unreliable. We therefore restrict our analysis to large-scale regional trends on the surface of Hartley 2's nucleus, which are unlikely to be the result of errors in the shape model.

The unstable scarp on the large lobe is not associated with activity on its face, although it does have a dust jet at its base (Bruck Syal et al. 2013). This suggests that this region did not recently experience avalanches that exposed interior ices preceding the flyby or during previous episodes of fast rotation. However, the instability of this region at all studied spin periods implies that any non-cohesive regolith would slide off of this surface, keeping underlying cohesive materials exposed at the surface. While this feature of the nucleus could result from a non-principal axis rotation that induced avalanches to build up a topographic feature on the nucleus, such rapid non-principal axis rotation states are unlikely (as previously noted). This scarp region may therefore be an outcrop of cohesive materials that generally underlie the cometary regolith, similar to the Hathor terrain on comet Churyumov-Gerasimenko (67P) (Sierks et al. 2015; Thomas et al. 2015). However, its lack of observed $CO_2$ activity suggests that it has long been exposed at the surface, allowing $CO_2$ ices to recede below the diurnal thermal skin depth, and possibly the orbital thermal skin depth.

Although the principal rotation period of Hartley 2's nucleus was increasing by ~1 minute/day during the DIXI encounter (Drahus et al. 2011, Belton et al. 2013), this rate is not maintained during an entire orbit (Knight et al. 2015). Drahus et al. (2013) estimated the orbitally averaged decrease in the principal axis rotation frequency to be ~0.012 $hr^{-1}$ per orbit. At this rate, Hartley 2 would have been spinning fast enough (period of 11 hours) to induce surface slope instabilities ~3-4 orbits prior to the DIXI encounter. This suggests that Hartley 2 was reactivated between 1984 and 1991. Interestingly, this coincides with the discovery of the comet by Malcolm Hartley in 1986, and activation of the small lobe of the nucleus would have certainly facilitated the discovery of the comet. Such a recent activation is consistent with systematic observations of Hartley 2 that began during its 1991 apparition, which suggest the activity of the nucleus has been diminishing over time (Meech et al. 2011; Knight & Schleicher, 2013), as would be expected as the exposed surface of the nucleus evolves thermophysically. It is therefore quite likely that Hartley 2 was a relatively dormant comet that reactivated shortly before its discovery, possibly during its 1985 perihelion passage.

During the rotationally-induced avalanches, material flows downhill toward the tip of the small lobe. If the $CO_2$ ices of Hartley 2 were located at a similar depth as they are on Tempel 1 (9P), then the thickness of the avalanching material is equal to the orbital thermal skin depth of Tempel 1's seasonally-active $CO_2$ (Feaga et al. 2007), which we estimate to be up to ~10-100 cm by scaling Tempel 1's ~1-10 cm diurnal skin depth (Groussin et al. 2007; Davidsson et al. 2013) to its orbital period. Hartley 2 has a surface area of 5.24 $km^2$ (Thomas et al. 2013), ~16% of which is sublimating $CO_2$ (Samarasinha & Mueller, 2013). If we assume that all of this $CO_2$ activity is located on the small lobe, then surface material covering ~0.8 $km^2$ (~2.4x$10^{10}$ kg) would avalanche toward the tip of the small lobe, severely restricting cometary activity.

However, since DIXI observed this region to be highly active (A'Hearn et al. 2011), this material needs to be removed from the surface.

At a spin period of 11 hours, the local normal acceleration (gravity plus rotation) at the tip of the small lobe is ~$5 \times 10^{-6}$ m/s$^2$, which is a factor of 5 smaller than during the DIXI flyby (due to the faster spin rate). Thus, it would have been easier to remove material from the surface through gas drag during periods of fast rotation. If we assume that the $CO_2$ ices, once exposed to sunlight during the avalanches, quickly warmed up and reached a sublimative equilibrium, then the sublimating $CO_2$ would exert a dynamic sublimation pressure between 0.001 Pa (at aphelion) and 0.1 Pa (at perihelion) (Steckloff et al. 2015; Steckloff & Jacobson, 2016) on the surface materials. This results in a sublimative force that exceeds the local normal force (gravity plus centripetal) at the tip of the small lobe for spherical chunks of material less than ~0.5-50 m in diameter, assuming a density equal to that of the bulk nucleus density. Thus, chunks smaller than ~0.5 - 50 m could be lofted off the surface of the small lobe during the avalanche. This is greater than the expected ~0.1-1 m depth to the $CO_2$ sublimation front (Groussin et al. 2007; Davidsson et al. 2013). Thus, if the small lobe experienced a series of avalanches, avalanche debris would be unlikely to cover and shut down the activity of exposed downslope $CO_2$ ices, allowing an area of $CO_2$ activity to grow through successive avalanches.

Additionally, avalanche debris accelerates as it travels downslope. We computed the net specific potential (gravitational plus rotational potential per unit mass) of Hartley 2's surface at the time of the avalanche to determine the maximum kinetic energy, and therefore maximum speed, that the avalanche debris could acquire (see Figure 6). We find that materials that originate from the edge of the small lobe's active area could have a specific kinetic energy of up to ~4 mJ/kg, and reach a maximum speed of nearly 0.1 m/s, which is below the nucleus' escape velocity but fast enough to slide off the tip of the small lobe. As a result, large pieces of material can leave the small lobe and fall back onto the nucleus, and may be the source of the ~20-40 m mounds and rough surface terrains observed by the DIXI mission (Thomas et al. 2013).

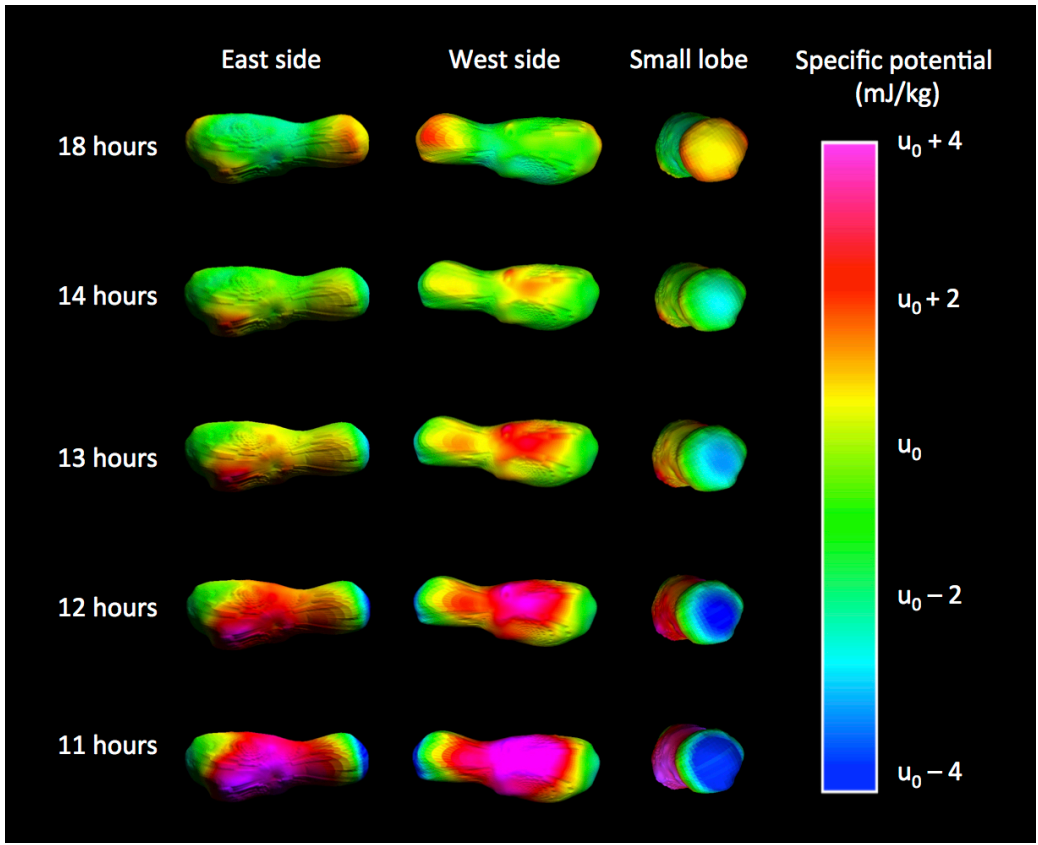

**Figure 6:** *Specific Potential Map of Hartley 2 at Various Spin Period.* We compute the net specific potential (gravitational plus rotational potential per unit mass) for various rotation periods. Because we only care about potential differences we plot each case with a color bar that spans 8 mJ/kg, roughly centered about the median specific potential of the surface ($u_0$). We clearly see that the regions of low specific potential migrate over the surface of the nucleus as the spin period increases. At the time of avalanche that excavated the small lobe, the potential lows were located at the tips of the lobes, migrating to waist and central mound of the nucleus by the time of the DIXI flyby.

We replotted the Thomas et al. (2013) binned size distribution data of these mounds on the surface of Hartley 2, and found that their cumulative size-frequency distribution (SFD) has a power-law index of -4.1. This is consistent with the range of indices of -5.6 - -3.7 for the differential SFD of the chunks of material found in Hartley 2's inner coma (Kelley et al. 2013) and with the index of -4.0±0.3 from *WISE/NEOWISE* observations of Hartley 2 (Bauer et al. 2011), suggesting that an ejection and fallback origin of these mounds is plausible. (See Figure 7). In any case, it is quite likely that the avalanching surface materials would leave the surface of the region of $CO_2$-driven activity on the small lobe, and may end up being deposited on the rest of the nucleus.

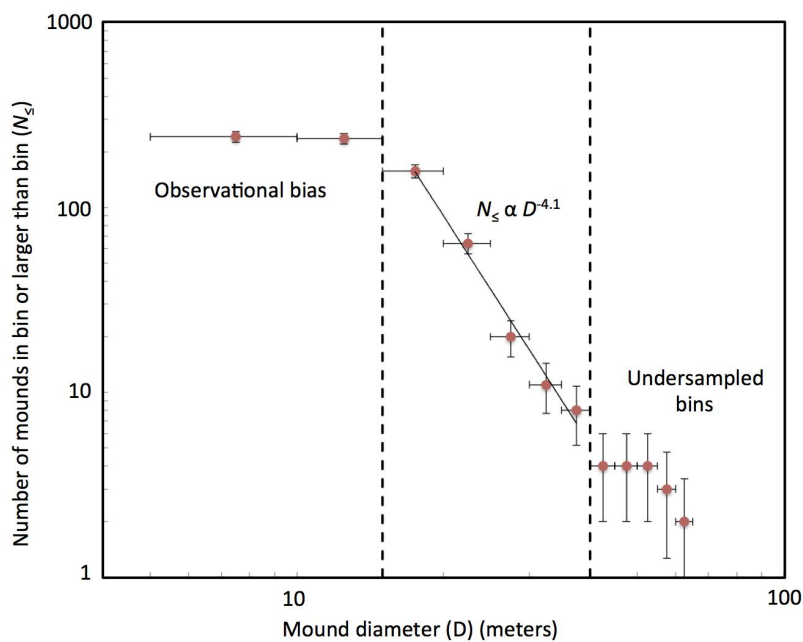

**Figure 7:** *Size-Frequency Distribution of Mounds on Nucleus of Hartley 2.* We replotted the binned size-frequency data of mounds on the surface of Hartley 2 from Thomas et al. (2013) to compute their cumulative size frequency distribution (cumulative SFD). Thomas et al. (2013) placed the data in 5 meter bins (the source of the horizontal error bars). Vertical error bars are standard square-root of *N*. Because data is binned, large bins become undersampled due to low probability of mound having a large size. We exclude these bins, which would otherwise skew the power-law fit to the SFD. We exclude the smallest-sized bins, which are near the resolution of the Deep Impact MRI, and therefore exhibit observational bias. We find that the mounds follow

a power-law cumulative SFD, with a power-law index of -4.1, consistent with the SFD of icy chunks in the inner coma of Hartley 2 during the DIXI flyby (Kelley et al. 2013)

We also computed the net specific potential (gravitational plus rotational potential per unit mass) of Hartley 2's surface at other spin periods. Since materials are preferentially deposited in potential minima, we consider the location of the potential minima at the time of Hartley 2's avalanche to determine where the materials are preferentially deposited. We find that the potential minima are located at the tips of each lobe. However, since the avalanche activated $CO_2$ activity would prevent this material from settling back onto the surface of the small lobe, the avalanche debris is most likely to settle at the tip of the large lobe. Thus, our model of Hartley 2's activation predicts that a rough, hummocky avalanche debris deposit should be located at the tip of the large lobe, consistent with observations of hummocky terrain near this location (see Figure 1). Furthermore, our model predicts the existence of other depositional terrains as the location of potential lows migrates about the nucleus during the lengthening of the principal rotation period of the nucleus from ~11 hours at the time of the avalanche to the 18.34 hour period during the DIXI flyby (see Figure 6).

At spin periods longer than ~14 hours, the potential low of the nucleus migrates from the tips of the lobes to the large region on the center of the imaged portion of the large lobe (denoted by Bruck-Syal et al. [2013] as a "central mound"). During this episode of intermediate spin periods, the surface of the nucleus is stable (slope angles below the angle of repose), inhibiting the formation of large-scale hummocky avalanche deposits. However, the $CO_2$-driven activity at the tip of the small lobe is ejecting grains composed of $H_2O$ ice and dust (A'Hearn et al. 2011), some of which settle back onto the surface. As these grains warm up, the $H_2O$ ice sublimates, forming a fluidized flow (Belton & Melosh, 2009) that settles in the potential well. Thus, our avalanche theory predicts the formation of a smooth terrain that approximates an equipotential surface on the central mound, also consistent with observations. At a spin period longer than ~16 hours, the waist region of the nucleus joins this potential low, where fluidized icy grains should flow and settle, forming a smooth, equipotential surface (A'Hearn et al. 2011; Thomas et al. 2013; Richardson & Bowling, 2014), consistent with observations. Finally, we note that the unstable scarp on the large lobe is part of a region that is a persistent location of high potential at all studied spin periods, consistent with this region being a cohesive outcrop of underlying materials.

From this sequence of events, we can place relative ages on most of the imaged surface of Hartley 2. The region of persistent potential high is likely the oldest terrain on the surface, with the unstable scarp representing an outcrop of the nucleus on the surface. The $CO_2$-driven active region and hummocky terrain at the tip of the large lobe, which formed simultaneously ~1984-1991 are the oldest terrains formed since the avalanche. The smooth terrain of the central mound formed next, when the rotation period was longer than ~14 hours. This terrain developed over time, as fallback material accumulated on the surface of the nucleus and fluidized. Finally, the smooth waist region joined the potential low at spin periods longer than ~16 hours, and formed most recently as fluidized fallback material settled.

It is likely that surface slope instabilities across the small lobe did not set in simultaneously. More likely, a series of smaller avalanches excavated buried $CO_2$ ices, rather than one large avalanche, causing a more gradual brightening of the comet (gradual, long-lived outburst). As this outburst faded, the comet would dim, but would settle to a much brighter magnitude than before the outburst due to the activation of $CO_2$-driven activity on the small lobe. The $CO_2$-driven regions would resist being covered by an avalanche because their activity would be able to loft this material off the surface, preventing it from settling.

At the ~11 hour rotation period at which surface slope instabilities set in, the waist of the nucleus enters a tensile state. Although only a small amount of cohesion (~2.8 Pa) is needed to hold the nucleus together, the waist may still experience a slow creep of the nucleus material, which may lead to an elongation of the neck, and further enhance the bilobate shape of the nucleus. The rest of the nucleus remains in compression, with most of the large and small lobes experiencing very small stresses (~1 Pa) at all rotation periods between 11 and 18 hours. This lack of significant compression or change in stress state would allow the nucleus to remain uncompacted, and may explain why the nucleus of comet Hartley 2 appears to be significantly less dense than other observed comet nuclei (Richardson et al. 2007; Thomas et al. 2013; Sierks et al. 2015).

In proposing that the activity of Hartley 2 is the result of rotationally-induced avalanches, we require a mechanism to spin up the nucleus to the onset of surface-slope instabilities. Although we suggest that Hartley 2 was relatively dormant prior to reactivation, it may still have exhibited some activity, which can generate sublimative torques. Because our results suggest that the nucleus spun down from an ~11 hour period in ~30 years, it is reasonable to assume that comparable levels of sublimative activity could spin the nucleus up to the point of inducing surface slope instabilities on the same timescale. However, if we linearly extrapolate this ~30 year timescale to the typical ~3 x $10^5$ year dynamical lifetime of a Jupiter Family Comet (Duncan et al. 2004), we find that Hartley 2 only requires an average of ~0.01% of its modern activity level to generate surface slope instabilities that induce avalanches and reactivate the nucleus over a typical JFC dynamical lifetime. Such a low level of activity may be undetectable. If the brightness of the coma scales linearly with activity, then a four order-of-magnitude drop in activity would correspond to a drop in the brightness of the coma of 10 magnitudes. It is therefore plausible that undectable, very weak activity could have spun up the nucleus of a dormant Hartley 2 and activate the activity on the small lobe.

Our work suggests that comet Hartley 2 was recently a dormant comet, a class of comets that recent work suggests exists (Kresak, 1987; Levison et al. 2006). While dormancy may be a common phase of Jupiter Family Comets, their mechanisms of reactivation are poorly understood. Bruck Syal et al. (2013) show that a rapid release of energy into the subsurface of comet Hartley 2 can trigger the formation of well-collimated jets. However, whereas Bruck Syal et al. (2013) propose that amorphous ice crystallization could provide the energy required to initiate jet formation, amorphous ice has yet to be detected on a comet (Huebner, 2009; Lisse et al. 2013), and another explosive source of heat on comet nuclei has not been identified. We have shown that rotationally-induced avalanches can excavate ices and reactivate a comet.

Conclusions

We have shown that the distribution of the activity of comet 103P/Hartley 2, which is driven by diurnally controlled carbon dioxide sublimation, is consistent with rotationally induced surface slope instabilities from a recent episode of fast rotation. At a rotation period of ~11 (10-13) hours, the region of Hartley 2's surface that is bounded by its carbon dioxide activity becomes unstable and avalanches toward the tip of the small lobe, excavating buried materials beneath. These avalanched materials would likely be lofted off the surface of the nucleus at less than escape velocity and be redeposited on other parts of the nucleus, forming hummocky terrains. We further show that spinning up the nucleus from its DIXI encounter spin period of 18.34 hours to this faster spin period does not induce surface slope instabilities on any other part of the nucleus prior to generating instabilities at the tip of the small lobe. Additionally, we have shown that the stresses within the nucleus are small enough in magnitude that the shape of the nucleus can be maintained if it possesses a bulk cohesive strength of at least 2.8 (2.0-4.0) Pa, which is consistent with strength estimates of other comet nuclei (Sekanina & Yeomans, 1985; Asphaug & Benz, 1996; Melosh, 2011; Bowling et al. 2015; Groussin et al. 2015; Thomas et al. 2015; Steckloff et al. 2015). We therefore conclude that rotational spin up is responsible for either initiating or maintaining the diurnally-controlled carbon dioxide driven activity of comet Hartley 2, and may have formed the observed knobby, hummocky terrains in the process. Our model of activating Hartley 2 predicts the formation of smooth fluidized terrains on the central mound and waist as the nucleus rotation period lengthened to the 18.34 hours of the DIXI flyby, consistent with observations. We discuss how this mechanism can potentially reactivate dormant comet nuclei, and how the shape of Hartley 2 may be responsible for its unusually low density by limiting the magnitude of the stresses that compact of the nucleus.

Acknowledgements

The authors would like to thank Mike A'Hearn and an anonymous reviewer for their helpful suggestions and comments that greatly improved this manuscript. The authors would also like to acknowledge Patrick Taylor of Arecibo Observatory and Hal Levison of Southwest Research Institute for helpful conversations that improved this manuscript.

References

A'Hearn et al., Deep Impact: Excavating Comet Tempel 1. Science 310, 258-264 (2005)

A'Hearn et al., EPOXI at Comet Hartley 2, Science 332,1396 (2011)

Asphaug, E. & Benz, W. Size, density, and structure of Comet Shoemaker-Levy 9 inferred from the physics of tidal breakup. Icarus 121, 225-248 (1996)

Bauer, J.M. et al., 2011. WISE/NEOWISE observations of Comet 103P/Hartley 2.

Astrophys. J. 738, 171.

Belton, M.J.S. & Melosh, H.J.; Fluidization and multiphase transport of particulate cometary material as an explanation of the smooth terrains and repetitive outbursts on 9P/Tempel 1. Icarus 200, 280-291 (2009)

Belton, M.J.S. et al.; The complex spin state of 103P/Hartley 2: kinematics orientation in space. Icarus 222, 595-609 (2013)

Blum, J. et al.; Comets formed in solar-nebula instabilities! – An experimentatl and modeling attempt to relate the activity of comets to their formation process. Icarus 235, 156-169 (2014)

Bowling, T.B.; Steckloff, J.K.; Melosh, H.J.; Graves, K.J.; The Strength of Comet 67P/Churyumov-Gerasimenko. *American Astronomical Society – Division of Planetary Science Meeting #46*, Abstract #100.03 (2014)

Bruck Syal, M.; Schultz, P.H.; Sunshine, J.M.; A'Hearn, M.F.; Farnham, T.L.; Dearborn, D.S.P; Geologic control of jet formation on COmet 103P/Hartley 2. Icarus 222, 610-624 (2013)

Davidsson, B.J.R. et al.; Thermal inertial and surface roughness of Comet 9P/Tempel 1. Icarus 224, 154-171 (2013)

Drahus, M. et al.; Rotation state of comet 103P/Hartley 2 from radio spectroscopy at 1 mm. Ap. J. L. 734:L4 (6pp) (2011)

Duncan, M.; Levison, H.; Dones, L.; "Dynamical Evolution of Ecliptic Comets" in Comets II Editors: Festou, M, Keller, H.U., Weaver, H.A.; University of Arizona Press-Tucson, AZ. p. 193-204 (2004)

Feaga, L.; Sunshine, J.; Protopapa, S.; A'Hearn, M.; Farnham, T.; Kelley, M.; Besse, S.; Groussin, O.; Comet 103P/Hartley's volatiles within 100 kilometers: Sources of water and volatile dependence on illumination. *Asteroids, Comets, Meteors 2014*, 157 (2014)

Groussin, O. et al.; Surface temperature of the nucleus of Comet 9P/Tempel 1. Icarus 187, 16-25 (2007)

Groussin, O. et al. ; Gravitational slopes, geomorphology, and material strengths of the nucleus of comet 67P/Churyumov-Gerasimenko from OSIRIS observations. A & A, in press. (2015)

Gulkis, S. et al.; Subsurface properties and early activity of comet 67P/Churyumov-Gerasimenko. Science 347, (2015) doi: 10.1126/science.aaa0709

Hässig, M. et al; Time variability and heterogeneity in the coma of 67P/CHuryumov-Gerasimenko. Science 347, (2015) doi: 10.1126/science.aaa0276

Harmon, J.K. et al.; Radar observations of 8P/Tuttle: A contact-binary comet. Icarus 207, 499-502 (2010)

Hirabayashi, M. & Scheeres, D.J., Stress and failure analysis of rapidly rotating asteroid (29075) 1950 DA. Ap. J. L. 798.1, L8 (2015)

Huebner, W.F. Origins of cometary materials. In: Origin and Early Evolution of Comet Nuclei, p. 5 - 25. Editors: Balsiger, H., Altwegg, K., Huebner, W., Owen, T., Schulz, R. (2009)

Keller, H.U. et al.; First Halley Multicolour camera imaging results from Giotto. Nature 321, 320-326 (1986)

Kelley, M.S. et al. A distribution of large particles in the coma of Comet 103P/Hartley 2. Icarus 222, 634-652 (2013)

Knight, M.M. & Schleicher, D.G.; CN Morphology studies of comet 103P/Hartley 2; AJ 141, 183 (2011)

Knight, M.M. & Schleicher, D.G.; The highly unusual outgassing of Comet 103P/Hartley 2 from narrowband photometry and imaging of the coma. Icarus 222, 691-706 (2013)

Knight, M.M.; Bueller, B.E.A.; Samarasinha, N.H.; Schleicher, D.G.; A Further Investigation of Apparent Periodicities and the Rotational State of Comet 103P/Hartley 2 from Combined Coma Morphology and Light Curve Data Sets. AJ 150:22, 14 pp. (2015)

Kresák, L.; Dormant phases in the aging of periodic comets.; A & A 187, 906-908 (1987)

Lambe, T.W. & Whitman, R.v.; Soil Mechanics. John Wiley & Sons: New York. (1969)

Levison, H.F. et al.; ON the origin of the unusual orbit of Comet 2P/Encke. Icarus 182, 161-168 (2006)

Lisse, C.; Bar-Nun, A.; Laufer, D.; Belton, M.; Harris, W.; Hsieh, H.; Jewitt, D.; Cometary Ices. In *The Science of Solar System Ices*, 455 – 485. Editors: Gudipati, M.S. & Castillo-Rogez, J.; Springer Publishing: New York. (2013)

Meech, K.J. et al.; EPOXI: Comet 103P/Hartley 2 observations from a worldwide campaign; Ap. J. L. 734:L1 (9pp) (2011)

Melosh, H.J.; Slopes and mass movement. In *Planetary Surface Processes* Edn. 1; Cambridge University Press: Cambridge, UK. p326-345 (2011)

Oberst, J. et al.; The nucleus of Comet Borrelly: a study of morphology and surface brightness. Icarus 167, 70-79 (2004)

Pravec, P. & Harris, A.W.; Fast and slow rotation of asteroids. Icarus 148, 12-20 (2000)

Richardson, J.E.; Melosh, H.J.; Lisse, C.M.; Carcich, B.; A ballistics analysis of the Deep Impact ejecta plume: Determining Comet Tempel 1's gravity, mass, and density. Icarus 190, 357-390 (2007)

Richardson, J.E. & Melosh, H.J.; An examination of the Deep Impact collision site on Comet Tempel 1 via Stardust-NExT: Placing further constraints on cometary surface properties. *Icarus* **222**, 492-501 (2013)

Richardson, J.E. & Bowling, T.J., Investigating the combined effect of shape, density, and rotation on small body surface slopes and erosion rates., Icarus 234, 53-65 (2014)

Samarasinha, N. H. & Mueller, B.E.A. Relating changes in cometary rotation to activity: Current status and applications to comet C/2012 S1 (ISON). Ap. J. L. 775:L10 (2013)

Schultz, P.H.; Hermalyn B.; Veverka, J.; The Deep Impact crater on 9P/Tempel-1 from Stardust-NExT. *Icarus* **222**, 502-515 (2013)

Sekanina, Z.; Sublimation rates of carbon monoxide and carbon dioxide from comets at large heliocentric distances. *Asteroids, Comets, Meteors 1991*, 545-548 (1992)

Sekanina, Z. & Yeomans, D.K.; orbital motion, nucleus precession, and splitting of periodic comet Brooks 2. *Astron. J.* **90**, 2335-2352 (1985)

Sierks, H. et al.; On the nucleus structure and activity of comet 67P/Churyumov-Gerasimenko. Science 347 (2015) DOI:10.1126/science.aaa1044

Steckloff, J.K. et al., Dynamic sublimation pressure and the catastrophic breakup of Comet ISON. Icarus 258, 430-437 (2015)

Steckloff, J.K. & Jacobson, S.A.; The formation of striae within cometary dust tails by a sublimation-driven YORP-like effect. Icarus 264, 160-171 (2016)

Thomas, P.C. et al. Shape, density, and geology of the nucleus of comet 103P/Hartley 2. Icarus 222, 550-558 (2013)

Thomas, N. and 58 coauthors; The morphological diversity of comet 67P/Churyumov-Gerasimenko. *Science* **347**, (2015) DOI: 10.1126/science.aaa0440

Werner, R.A., The gravitational potential of a homogeneous polyhedron or don't cut corners. Celest. Mech. Dynam. Astron. 59, 253-278 (1994)